\documentstyle[prl,aps,epsfig]{revtex}

\twocolumn
\narrowtext

\begin{document}
\draft
\wideabs{
\title{Counting Cold Collisions}
\author{B.\ Ueberholz, S.\ Kuhr, D.\ Frese, D.\ Meschede, and V.\ Gomer}
\address{Institut f\"ur Angewandte Physik, Universit\"at Bonn \\
         Wegelerstr.\ 8, D-53115 Bonn, Germany}
\maketitle
\begin{abstract}
 We have experimentally explored a novel possibility to study exoergic
 cold atomic collisions. Trapping of small countable atom numbers in a
 shallow magneto-optical trap and monitoring of their temporal dynamics
 allows us to directly observe isolated two-body atomic collisions and
 provides detailed information on loss statistics. A substantial fraction
 of such cold collisional events has been found to result in the loss of
 one atom only. We have also observed for the first time a strong optical
 suppression of ground-state hyperfine-changing collisions in the trap
 by its repump laser field.
\end{abstract}

 }

\begin{figure}[htb]
 \center{\epsfig{file=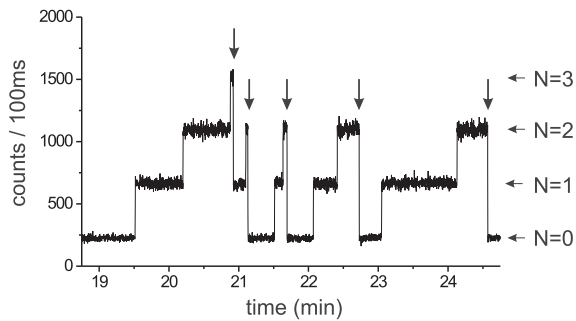,height=1.8in}}\vspace*{.4cm}
    \caption{Excerpt from typical MOT fluorescence signal
    observed with an avalanche photodiode.
    Five isolated cold collisions (two-atom losses) [6]
    are shown (arrows). The fluorescence level at atom number $N$ = 0
    corresponds to residual stray light. Clearly separated fluorescence
    steps can be easily resolved for atom number up to $N=20$.}
 \label{fig:fig1}
 \end{figure}

 The observation of physical phenomena at the atomic level often provides
 new insights into the details of the processes under study, usually hidden
 in ensemble samples. Trapping of individual neutral atoms in a
 magneto-optical trap (MOT) \cite{Raab87} realized in
 \cite{Hu94,Haubrich 96} allowed us to
 obtain information on internal and external atomic dynamics in the trap with
 excellent contrast \cite{wir}. Cold inelastic collisions
 are usually associated with high atomic densities and as a consequence
 with experiments on large numbers of trapped atoms.
 Here we study them with only a few atoms.
 In this situation one is able to monitor the instantaneous number of trapped
 atoms exactly and observe isolated load and loss events.
 In Fig.\ref{fig:fig1} we show an example of the dynamics of the trapped atom
 number in the MOT operating at constant conditions. Such a 'digitized' signal
 provides detailed information on collisional statistics we will analyze in
 the present Letter.
%
%

 By far the most popular choice for collision studies is the MOT providing
 dense samples of cold atoms. An exoergic collision converts internal
 atomic energy to kinetic energy equally divided between the colliding
 partners. If the transferred kinetic energy is greater than the recapture
 ability of the trap, the collision
 leads to trap loss. Here we experiment with a very shallow trap which is
 sensitive to low-energy collisional processes. It enables us to observe a
 strong optical suppression of ground-state hyperfine-changing collisions
 by the MOT repump laser field and to infer the corresponding rate constant.

 There has been extensive progress in both experimental and theoretical
 investigations of collisions between laser-cooled atoms \cite{rev}. The main
 method of observing collisions used so far has been to abruptly change
 experimental parameters (usually switching off or on an atomic beam loading
 the trap) and to watch the trap population decay or increase. The rate
 equation for the total number $N$ of trapped atoms is given by
 $\dot{N} = R -N/\tau_{\rm coll} - \beta \int n^2({\bf r},t)d^3r $,
 where the first two terms on the right hand describe the loading rate $R$
 and the loss rate due to collisions with background molecules, respectively.
 The last term represents intratrap collisional loss rate due to collisions
 between trapped atoms with a density profile $n({\bf r},t)$. In our
 experiments the number of trapped atoms is small and can be
 determined exactly. Thus information can be obtained from temporally resolved
 fluctuations in the atom number under conditions of dynamical equilibrium
 providing a possibility to work under constant experimental parameters.

\begin{figure}[hbt]
 \center{\epsfig{file=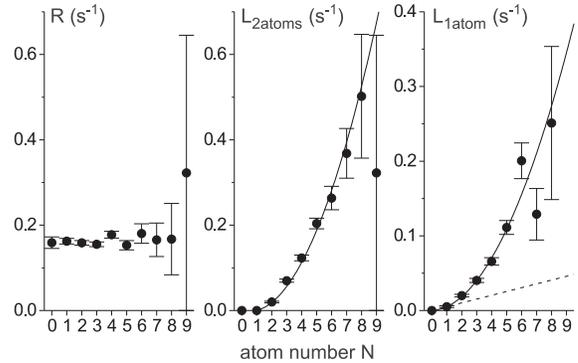,height=2.1in}}\vspace*{.4cm}
    \caption{Load and loss event rates
    as a function of the atom number $N$ measured for total cooling laser
    intensity $I=42$ mW/cm$^2$ and detuning $\delta=-3.35 \Gamma$
    and average atom number during the measurement $\langle N \rangle =2.6$ .
    Solid lines represent a second order polynomial
    fit. The dashed line shows the linear dependence due to collisions with
    background gas. Error bars indicate statistical error of load and loss
    occurrencies.}
 \label{fig:fig2}
 \end{figure}

 The number of trapped atoms and its temporal fluctuations is determined by
 balance between loading the MOT from the low-pressure atomic vapor and
 different loss mechanisms removing one or two atoms from the trap with the
 corresponding loss event rate
 $L_{\rm 1atom}(N)$ and $L_{\rm 2atoms}(N)$ at atom number $N$, respectively
 \begin{equation}
 \dot{N} = R- L_{\rm 1atom}(N)- 2L_{\rm 2atoms}(N) . \label{eq:rateold}
 \end{equation}
 We can distinguish various events with 100\% contrast
 and thus all terms on the right hand can be now measured
 independently. The two-atom losses obviously arise from collisions of trapped
 atoms $2L_{\rm 2atoms}(N)=\beta N(N-1)/V$, where $V=(\pi/2)^{3/2}r_0^3$ is
 the effective trapping volume.
 A simple guess is also to assume that $L_{\rm 1atom}$ in (\ref{eq:rateold}) is
 associated with collisions with background molecules
 $L_{\rm 1atom}(N)= N/\tau_{\rm coll}$. In order to check this assumption
 we have plotted in Fig.\ref{fig:fig2} the $N$-dependence of $R$, $L_{\rm 1atom}$
 and $L_{\rm 2atoms}$ by counting the corresponding load and loss events for each
 atom number at constant conditions. As expected we have found $R$ independent
 of $N$ and the two-atom loss rate scaling quadratically with the atom number.
 However and surprisingly a substantial part of the one-atom loss rate scales
 also quadratically with $N$ indicating 'soft' two-body collisions resulting
 in loss of only one atom \cite{check}. So we can rewrite (\ref{eq:rateold})
 with $L_{\rm 1atom}(N)= N/\tau_{\rm coll} + \beta_{\rm 1atom}N(N-1)/V$
 and $2L_{\rm 2atoms}(N)=\beta_{\rm 2atoms}N(N-1)/V$ in a form, that
 \begin{equation}
 \dot{N} = R - N/\tau_{\rm coll} - \beta N(N-1)/V,
 \end{equation}
 with $\beta= \beta_{\rm 2atoms}+\beta_{\rm 1atom}$. Both loss coefficients can be
 determined with good accuracy by quadratic fitting of the corresponding
 $N$-dependences.
%
%

 Our experimental setup has been described in detail elsewhere
 \cite{Haubrich 96,wir}. A six-laser-beam $\sigma^+$-$\sigma^-$ MOT is loaded
 from a low-pressure cesium vapor (at base pressure of better than
 10$^{-10}$ mbar). The trap laser detuning $\delta$ from the Cs cooling
 transition $F=4 \to F'=5$ (typically $\delta  = -3\Gamma$, in terms of the
 natural linewidth $\Gamma=2\pi\times 5.2$ MHz) is precisely controlled by a
 heterodyne phase-locking technique. To prevent the atom escaping from the
 cooling cycle by a decay into the $F=3$ ground-state, a second (repump) laser
 is introduced, resonant with the $F=3 \to F'=4$ transition and stabilized by
 standard techniques to $\Gamma/4$. The MOT magnetic quadrupole field is
 produced by permanent magnet discs with tunable field gradients up to
 $B'=dB_z/dz=800$ G/cm. Stiff magnetic field gradients lead to a strong
 reduction of the capture rate \cite{Hoepe 93}. It was also found that the MOT
 spring constant is linearly dependent on the quadrupole magnetic field
 gradient, while the temperature of trapped atoms is comparable to a low-field
 MOT \cite{Hoepe 93,Willems 97}.
 Thus the atomic density is expected to be proportional to $(B')^{3/2}$ and as
 a result the probability for collisions between trapped atoms rises
 dramatically $\propto (B')^3$. For $B'=375$ G/cm in our experiment the
 two-atom loss probability is comparable with the probability for background
 collisions: a comfortable situation for a measurement. The spatial
 distribution of the MOT fluorescence measured by a CCD camera has a
 Gaussian distribution with 1/e$^2$ radius $r_0$ between 7 and 24 $\mu$m
 depending on the laser intensity. The trap size was observed to be
 independent of the atom number (up to $N=8$) insuring that radiation trapping
 effects \cite{Walker 90} can be ignored. Fluorescence of the atoms trapped
 in the MOT is observed with avalanche photodiodes in single photon counting
 mode. Typical photon counting rates are 3-20 kHz per atom depending on the
 trap laser detuning and intensity. The probability for eventual
 misinterpretations (for example two one-atom losses occuring simultaneously
 within our integration time of 100 ms and detected as a two-atom loss event)
 is below 1\% and can be neglected here.

\begin{figure}[hbt]
 \center{\epsfig{file=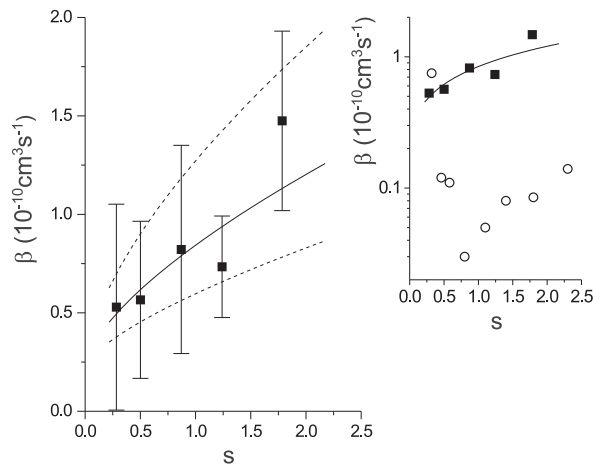,height=2.6in}}\vspace*{.4cm}
    \caption{ Total collisional loss coefficient $\beta$
    vs. cooling laser
    saturation parameter $s$ in our trap (squares). Solid line is a fit
    according to the semiclassical model (see text), dotted lines indicate
    uncertainty in the trap depth of factor two. Inset: same (log-) plot in
    comparison with standard Cs-MOT data (circles show an average over the
    scattered data in [11]).}
 \label{fig:fig3}
 \end{figure}

 Three main exoergic collisional processes in MOT's have been identified
 \cite{rev}: The fine-structure-changing collision (FCC) is represented by
 $A+A+\hbar \omega \to A^*_2(P_{3/2}) \to A^*(P_{1/2})+A+\Delta E_{FCC}$
 with the energy $\Delta E_{FCC}/2$ transferred to each atom. For Cs atoms
 $\Delta E_{FCC}/2k_B \approx 400$ K and FCC collisions ultimately cause an
 escape of both atoms from the MOT, usually no more than 1 K deep.
 For radiative escape (RE), spontaneous emission of a photon red-shifted from
 the atomic resonance takes place during the collision. The process is
 described by $A+A+\hbar \omega \to A^*_2 \to A+A+ \hbar \omega'$
 with energy $\hbar (\omega-\omega')/2$ transferred to each atom. The
 resulting kinetic energy is continuously distributed and the corresponding
 loss rate is sensitive to the effective trap depth. Exoergic
 hyperfine-changing collisions (HCC) on the molecular ground-state can also
 lead to losses if the trap is sufficiently shallow. For Cs, a change
 from 6s$^2$S$_{1/2} (F = 4)$ to 6s$^2$S$_{1/2}(F=3)$ in one of the colliding
 atoms transfers about $\Delta E_{HCC}/2k_B = 0.22$ K to each atom. For
 interpretation of trap-loss measurements it is essential to distinguish which
 collision processes are producing the trap loss. One such possibility is to
 use the intensity dependence of different loss channels. In a standard
 Cs MOT Sesko {\em et al.} \cite{Sesko 89} observed a rapid increase in
 $\beta$ as the intensity $I$ of trapping laser dropped below $s=0.8$, see
 Fig.\ref{fig:fig3} (data scaled to the effective saturation parameter
 $s$=$I/I_S/[1+(2\delta/\Gamma)^2]$). They interpreted this loss as due to
 hyperfine-changing collisions (HCC), but could not infer directly
 $\beta_{HCC}$ because even at the lowest MOT laser intensity the trap loss
 does not become independent of the trap depth. The rate constant
 $\beta_{HCC}$ has been estimated between
 10$^{-10}$ and 10$^{-11}$ cm$^3$s$^{-1}$. For higher intensities the trap
 becomes too deep for atoms to  escape after HCC and this contribution to
 the trap losses disappears. On the other hand the probability for one of
 the colliding atoms to be excited to an attractive
%
%
 potential curve
 (corresponding to the S$_{1/2}$+P$_{3/2}$ asymptote) increases with growing
 laser intensity and FCC and RE collisions constitute the main
 loss mechanism for usual traps. This interplay of different loss channels
 imparts a characteristic form to the intensity dependence of $\beta$ observed
 also in MOT's operating with other alkalis \cite{rev}.
 In our trap  we observe a similar loss rate coefficient at low excitation
 rates but the measured dependence in $s$ is quite different
 (Fig.\ref{fig:fig3}). Especially the steep rise in $\beta$ at low $s$ is
 absent.

 It can be easily explained by a substantially reduced 'trap depth' $\Delta U$
 (or more accurately by the maximum initial kinetic energy enabling an escape)
 due to large magnetic field gradients $B'$ in our MOT . The effective
 'deceleration distance' (where the scattering light force acting on the
 escaping atom is close to its maximum resonant value) scales as $1/B'$.
 Numerical simulations \cite{Kuhr 99} substantiate the interpretation that
 $\Delta U$ in Fig.\ref{fig:fig3} is a weak
 function of the trap laser intensity \cite{Ritchie 95} and is always
 smaller than the energy gained in a HCC. As a result every ground-state
 HCC process leads to a two-atom loss and the HCC contribution (about 50\% for
 low intensity in Fig.\ref{fig:fig3}) to the total trap losses is nearly
 independent of the trap laser intensity. In a shallow trap RE contribution
 to the total trap losses is also substantially larger in comparison to usual
 MOTs and is dominant at higher laser intensities. Based on \cite{Gallagher89}
 we have performed semiclassical numerical calculations of the distribution of
 the energy gained as a result of a RE collision for the parameters of our
 experiment giving reasonable agreement with our measurements (see
 Fig.\ref{fig:fig3}). This allows us to infer the value for the HCC
 collisional loss coefficient $2.0\cdot 10^{-11}$ cm$^{3}$s$^{-1}$ as
 intensity-independent offset in $\beta$. However, as we will see
 below, ground-state collisions can be strongly affected by the repump laser.

 As already mentioned a significant part (typically 10\% of the total loss
 rate) of two-body collisions in our trap leads to one-atom losses. The energy
 gained in a collision is equally divided between both atoms (the initial
 kinetic energy of colliding atoms of order of some 100 $\mu$K can surely be
 neglected). For producing an one-atom loss this energy must be comparable to
 the effective trap depth. Thus one-atom losses come obviously from RE
 processes. To be recaptured an escaping atom with initial kinetic energy of
 0.1 K has to scatter about 10$^3$ photons and thus statistical fluctuation of
 the trap depth is about 3\%. Statistical fluctuations also lead to deviations
 from originally counterpropagating straight-line paths
 of atoms leaving the trap which may be important because
 of the anisotropy of the trapping potential.
 As shown numerically \cite{Ritchie 95} and experimentally
 \cite{Kawanaka 93} $\Delta U$ varies by as much as a factor of 4 between
 the shallowest and deepest directions. It can be estimated \cite{Kuhr 99}
 that for RE collisions releasing more energy than 0.1 K about 50\% of
 collided atom pairs gain kinetic energy in the interval between 0.1 and
 0.4 K. But even under these assumptions the frequent occurrence of one-atom
 losses observed in the experiment can not be explained and needs some
 additional analysis \cite{Kuhr 99}.

\begin{figure}[tbh]
 \center{\epsfig{file=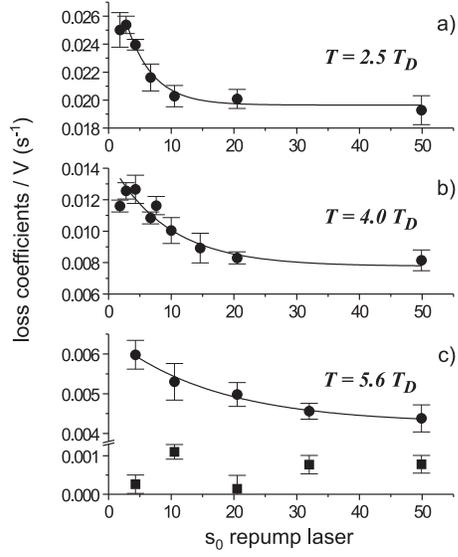,height=3in}}\vspace*{.4cm}
    \caption{
    Loss rates
    (circles: $\beta_{\rm 2atoms}/V$, squares: $\beta_{\rm 1atom}/V$)
    as a function of the repump laser intensity for different
    cooling laser parameters: a) s=0.87, b) s=1.74, c) s=5.19
    with representative values for $\beta_{\rm 1atom}/V$.
    Solid lines are calculated from Landau-Zener model (see text).
     }
 \label{fig:fig4}
 \end{figure}

 The 9 GHz blue detuning of the MOT repump laser is so large
 that it can not exert a force on a single escaping atom and thus can not effect
 the trap depth.
 However, we have observed a strong dependence of the loss rates on the
 repump laser intensity, Fig.\ref{fig:fig4}. We explain the exponential decay
 of the loss rates to a nearly constant value at high intensities by the
 process of optical shielding first reported in \cite{Bali92} and further
 investigated in \cite{Marcassa 94}.

 At small interatomic distance $R_C\approx100\mbox{ \AA}$ the repump laser is
 resonant with the repulsive quasimolecular potential corresponding to the
 asymptotic state $\left|S+P\right\rangle$. In the dressed-atom picture
 \cite{Bali92,Marcassa 94} the presence of the resonant laser field leads
 to an avoided crossing between the states $\left|S+S,n_\gamma \right\rangle$
 and $\left|S+P,n_\gamma-1\right\rangle$ with a Rabi splitting $\hbar \Omega$
 at the Condon point $R_C$. Here $n_\gamma$ is the photon number in the
 repump
 laser field. This splitting prevents the atom pairs from approaching in the
 $\left|S+S\right\rangle$ state close enough for the
 ground state collisions to occur. In our measurements the repump laser intensity
 is varied from $s_0=2$ to 50 producing values of $\hbar \Omega$ from
 $\hbar \Gamma$ to $7 \hbar \Gamma$. To reach $R_C$ the atom pairs must have
 a kinetic energy of $\hbar\Omega/2$, which is in the order of the Doppler
 temperature $k_B T_D=\hbar \Gamma/2$ ($T_D=125$ $\mu$K for Cs). Thus one expects a strong temperature
 dependence. Based on Landau-Zener model we have calculated the total
 suppression ratio by integration over the Maxwell-Boltzmann velocity
 distribution. In first approximation we find a simple scaling law
 for the probability of ground-state collisions with the repump
 laser intensity, $P_{HCC}\approx\exp{(-s_0/A(T))}$ \cite{Kuhr 99}. The decay
 constant $A(T)$ varys approximately as $A(T)=1+0.5(T/T_D)^2$.
 For simplicity we used only one repulsive molecular state with
%
%
 $V_{S+P}(R)=+C_3/R^3$ with $C_3=12$ a.u. \cite{Bussey 85}.
 However, we note that the result of our model changes very little
 if C$_3$ is changed by a factor of 2.
 The repump laser does not influence the light-induced collisions which appear
 therefore as an intensity independent offset. We see this in the RE events resulting in
 one-atom losses which are shown in Fig.\ref{fig:fig4}, c) as squares.

 To test our assumptions, we varied the MOT temperature
 by changing the cooling laser parameters (Fig.\ref{fig:fig4}) and found
 significantly different decay constants as expected,
 $A(T)$=$4.2\pm 0.8,9.2\pm 2.7,16.9\pm 4.2$ for a), b) and c), respectively.
 From the measured decay constants we can directly infer the temperatures $T$,
 see Fig.\ref{fig:fig4}. These results are in excellent agreement with
 previous temperature measurements \cite{wir,Hoepe 93}.
 Extrapolating the curves to zero repump intensity and taking into account the
 measured trap volumes we infer values for the total collisional rate
 coefficients $\beta_{HCC}$=$(3.3\pm 1.8), (4.6\pm 1.7), (4.3\pm 1.9) \cdot
 10^{-11}$ cm$^3$s$^{-1}$ for a), b) and c), respectively.
 Here the uncertainty in the trap size of $\Delta r_0 \approx 2$ $\mu$m
 for all measurements makes the main contribution to the errors in $\beta$
 (typical relative errors of non-normalized loss coefficients lie below 5\%
 for $\beta_{2atoms}/V$ and 20\%  for $\beta_{1atom}/V$ and about 50\% for
 absolute $\beta$ values).
 Note that although the absolute values for loss event rates
 and for atomic densities differ substantially for different temperatures
 in Fig.\ref{fig:fig4} all calculated $\beta_{HCC}$ are equal
 as expected for collisions between ground-state atoms. So we derive a final
 value for $\beta_{HCC} = (4.1\pm 1.0) \cdot 10^{-11}$ cm$^3$s$^{-1}$.
 The value of $\beta_{HCC}$ obtained from the measurement in Fig.\ref{fig:fig3}
 is now reasonably explained by optical suppression corresponding to
 repump intensity $s_0=4$ used for all data.

 In summary it has been shown that cold collision investigations on single
 trapped atoms provide novel and detailed information on the trap
 loss processes. We have measured the rate constant for the ground-state
 hyperfine-changing collisions partially hidden by strong optical shielding
 in previous studies. This intrinsic effect is always present in an alkali MOT.
 The method can also remove eventual ambiguities in experiments
 where an extra probe laser is introduced \cite{Sesko 89} in order
 to 'catalyse' different collisional loss channels. Note that such laser
 fields can also strongly affect the performance of the trap \cite{Feng 93}, and
 it is in general difficult to clearly discriminate between
 changed excitation conditions and changes in the atom number,
 both modifying the total fluorescence signal.
 It is also easy to generalize the method for studies of heteronuclear
 collisions \cite{Telles 99} where fluorescence from different species can
 easily be spectrally distinguished.
 One can furthermore speculate about possibility to watch the formation of
 an individual molecule from two atoms.

 We thank John Weiner for valuable discussions.
\bibliographystyle{prsty}

\begin{thebibliography}{99}
\bibitem{Raab87}
 E.L. Raab {\em et al.},
 Phys. Rev. Lett. {\bf 59}, 2631 (1987);

\bibitem{Hu94}
 Z. Hu and H.J. Kimble, Optics Lett. {\bf 19}, 1888 (1994);
 F. Ruschewitz {\em et al.}, Europhys. Lett. {\bf 34}, 651 (1996)

\bibitem{Haubrich 96}
 D. Haubrich {\em et al.},
 Europhys. Lett. {\bf 34}, 663 (1996)

\bibitem{wir}
 V. Gomer {\em et al.},
 Phys. Rev. A {\bf 58}, R1657 (1998);
 V. Gomer {\em et al.},
 Appl. Phys. B {\bf 67}, 689 (1998)

\bibitem{rev}
 P.S. Julienne, A.M. Smith and K. Burnett,
 in {\em Advances in Atomic, Molecular and Optical Physics},
 edited by D.R. Bates and B. Bederson (Academic Press, San Diego, 1993),
 {\bf 30}, 141;
 T. Walker and P. Feng, ibid.  {\bf 34}, 125 (1994);
 John Weiner, ibid.  {\bf 35}, 45 (1995);
 J. Weiner {\em et al.}, Rev. Mod. Phys. {\bf 71}, 1 (1999)

\bibitem{direct}
 first direct observation of isolated two-body atomic collisions in a MOT was
 reported in Ref. \cite{Willems 97}

\bibitem{check}
 We also checked that $R$ and $1/\tau_{\rm coll}$ increase with increasing Cs
 background pressure, while $\beta_{\rm 1atom}$ and $\beta_{\rm 2atoms}$ remain
 constant.

\bibitem{Hoepe 93}
 A. H\"ope {\em et al.},
 Europhys. Lett. {\bf 22}(9), 669 (1993);
 D. Haubrich {\em et al.},
 Optics Comm. {\bf 102}, 225 (1993)

\bibitem{Willems 97}
 P.A. Willems {\em et al.},
 Phys. Rev. Lett. {\bf 78}(9), 1660 (1997)

\bibitem{Walker 90}
 T. Walker {\em et al.},
 Phys. Rev. Lett. {\bf 64}, 408 (1990)

\bibitem{Sesko 89}
 D. Sesko {\em et al.},
 Phys. Rev. Lett. {\bf 63}, 961 (1989)

\bibitem{Kuhr 99}
 S. Kuhr {\em et al.},
 to be published

\bibitem{Ritchie 95}
 N.W.M. Ritchie {\em et al.},
 Phys. Rev. A {\bf 51}, R890 (1995)

\bibitem{Gallagher89}
 A. Gallagher and D.E. Pritchard, Phys. Rev. Lett. {\bf 63}, 957 (1989)

\bibitem{Kawanaka 93}
 J. Kawanaka {\em et al.},
 Phys. Rev. A {\bf 48}, R883 (1993);
 D. Wilkowski {\em et al.},
 Eur. Phys. J. D {\bf 2}, 157 (1998)

\bibitem{Bali92}
 S. Bali, D. Hoffmann, and T.Walker,
 Europhys. Lett. {\bf 27}, 273 (1992)

\bibitem{Marcassa 94}
 L. Marcassa {\em et al.},
 Phys. Rev. Lett. {\bf 73}, 1911 (1994);
 S.R. Muniz {\em et al.},
 Phys. Rev. A {\bf 55}, 4407 (1997)
 V. Sanchez-Villicana {\em et al.},
 Phys. Rev. Lett. {\bf 74}, 4619 (1995)

\bibitem{Bussey 85}
 B. Bussery {\em et al.},
 J. Chem. Phys. {\bf 82}, 3224 (1985)

\bibitem{Feng 93}
 P. Feng {\em et al.},
 Phys. Rev. A {\bf 47}, R3495 (1993)

\bibitem{Telles 99}
 G.D. Telles {\em et al.},
 Phys. Rev. A {\bf 59}, R23 (1999)

\end{thebibliography}

 \noindent

 \noindent

 \noindent

 \noindent

\end{document}